\documentclass[12pt]{report}
\usepackage{rotating}
\usepackage{floatpag}
\rotfloatpagestyle{empty}
\usepackage{amsthm}
\usepackage{graphicx}
\usepackage[bf]{subfigure}
\usepackage{color}
\usepackage{url}
\usepackage{amsmath}
\usepackage{amssymb}
\usepackage[T1]{fontenc}
\usepackage[bf]{subfigure}
\usepackage{rotating}
\usepackage{subfigure}
\usepackage{mathptmx}
\usepackage{caption}
\makeindex             

\usepackage{diagbox}

\usepackage[top=3cm, bottom=3cm, left=2.3cm, right=2cm]{geometry}

\begin{document}

\chapter*{Network centrality: an introduction}

\emph{Francisco Aparecido Rodrigues}\\
\emph{Instituto de Ci\^{e}ncias Matem\'{a}ticas e de Computa\c{c}\~{a}o,\\
Universidade de S\~{a}o Paulo - Campus de S\~{a}o Carlos,\\Caixa Postal 668, 13560-970 S\~{a}o Carlos, SP, Brazil.}\\
\emph{francisco@icmc.usp.br}

\vspace{1cm}

\noindent\textbf{Abstract}\\
\begin{small}
Centrality is a key property of complex networks that influences the behavior of dynamical processes, like synchronization and epidemic spreading, and can bring important information about the organization of complex systems, like our brain and society. There are many metrics to quantify the node centrality in networks. Here, we review the main centrality measures and discuss their main features and limitations. The influence of network centrality on epidemic spreading and synchronization is also pointed out in this chapter. Moreover, we present the application of centrality measures to understand the function of complex systems, including biological and cortical networks. Finally, we discuss some perspectives and challenges to generalize centrality measures for multilayer and temporal networks.
\end{small}

\section{Introduction}
\label{sec:1}

Complex systems are made up of connected elements whose interactions are nonlinear~\cite{BarYam97,Mitchell09}. Examples of complex systems include our society, the Internet, our brain and cellular interactions~\cite{Mitchell09}. The modeling of complex system behavior has attracted the attention of researchers from several areas, including Mathematics, Physics, Biology, Computer Science and Engineering~\cite{Costa011}. The study of these systems has several applications in Science and Technology, including the control of disease spreading~\cite{Pastor015} and adjustment of synchronization in data transmission~\cite{Pikovsky03}. Since the last century, researchers have verified that the structure of several complex systems can be modeled as complex networks~\cite{Boccaletti06, Costa07, Barabasi016}. These networks are characterized by a very heterogeneous organization, presenting a special set of highly connected elements, whereas the remainder of the components is low connected~\cite{Barabasi99}. This ubiquitous scale-free architecture is observed in numerous systems, from food webs to collaboration networks of scientists~\cite{Costa011}, and has important implication for system dynamics. For instance, previous works have verified that the epidemic threshold for disease propagation depends on the level of network heterogeneity, being close to zero in a scale-free network~\cite{Pastor015}. Thus, scale-free networks are the ideal medium for disease transmission, explaining the quick propagation of viruses in our society, like those that cause influenza~\cite{Keeling08}. Moreover, the network organization influences the emergence of synchronization in systems like power grids and our brain~\cite{Arenas08, Rodrigues016}. In this case, a set of self-sustained oscillators interacts adjusting their phases and after a given critical coupling strength, a collective behavior emerges. This critical coupling strength also depends on the network heterogeneity. Other dynamical processes, including cascade failures, percolation, and voter models, also depend on the network structure~\cite{Barrat08}.

Since the structure of complex networks is very heterogeneous, it is expected that some nodes are more important than others in some sense. This importance can be quantified by the network centrality, as we expected that the most central nodes are the most influential ones and can propagate their information content easier than other nodes~\cite{Kitsak010, Arruda014}. However, there is no general definition of centrality and, therefore, many centrality measures have been developed~\cite{Costa07}. In this chapter, we review the main centrality measures used to characterize complex networks structure and present their advantages and limitations. The influence of centrality on dynamical processes is also discussed in terms of epidemic spreading and synchronization. We show that central nodes are the most influential spreaders of diseases. However, the identification of these nodes through centrality measures depends on the network structure. In the case of synchronization, we show that when the natural frequency of Kuramoto oscillators is correlated with their number of connections, the system undergoes a first order phase transition~\cite{Gardenes011}, differently from the second order phase observed when the natural frequency is independent of network structure~\cite{Rodrigues016}. Applications of network centrality in some areas, including systems biology and neuroscience, are also presented here. Finally, we discuss some perspectives regarding the study of network centrality and generalizations to more intricate networks, including multilayer organization and temporal networks.

\section{Centrality measures}\label{Sec:centrality}

A complex network is a graph $G$ with particularly intricate structure, made of an ordered pair of disjoint sets $(V,E)$, where $V$ is a set of elements called vertices (nodes) and $E$ is a subset of ordered pairs of distinct elements of $V$, called edges or arcs. If the network is undirected, i.e.\ for every connection going from each pair $i$ to $j$ has a connection $j$ to $i$, the links are called edges.  Otherwise, directed connections are called arcs. Network edges can also have weights, e.g.\ indicating the strength of the interaction between two nodes. In this chapter, we consider only undirected and unweighted networks to discuss the centrality measures.

In complex networks, due to its highly heterogeneous structure, some nodes can be considered as more important than other ones. For instance, in social networks, some people, like celebrities and politicians have a lot of followers and can propagate information easier than ordinary subjects. Hence, these nodes can be considered as central. However, this definition of centrality is not unique, since we can define it in terms of the load that each node receives. For example, in a street network, an urban region is central whether it presents traffic jams and is more accessed than other places. Thus, the definition of centrality is not general and depends on the application. Since we do not have a consensus about the general definition of centrality, several measures have been proposed, where each one considers specific concepts. 

\begin{figure}[!t]
\begin{center}
\includegraphics[width=.9\linewidth]{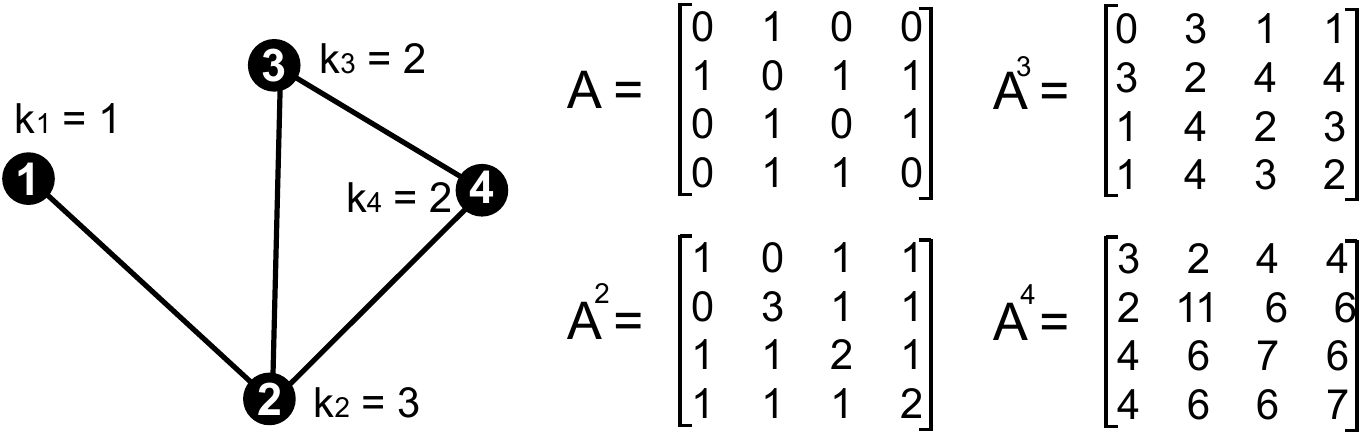}
\end{center}
\caption{An illustration of the concepts of degree and the power of the adjacency matrix $A$. The elements $(A_{ij})^n$ give the number of walks of length $n$ between nodes $i$ and $j$. The degree $k_i$ of each node is indicated in the network.}
\label{fig:degree-ex}
\end{figure}

\begin{figure}[!t]
\begin{center}
\subfigure[Degree]{ \includegraphics[width=.3\linewidth]{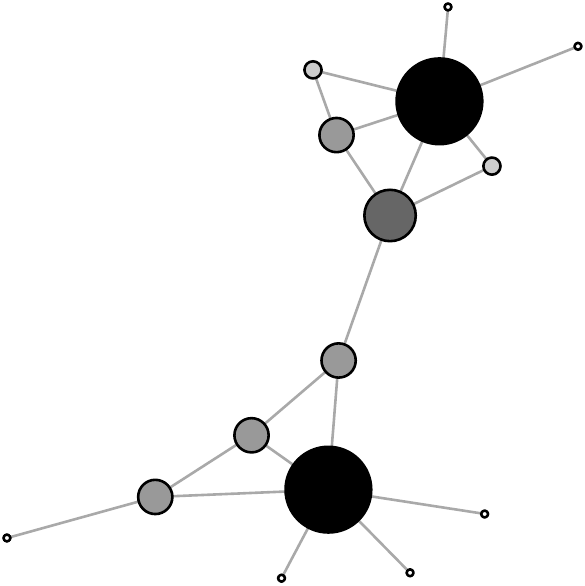}}
\subfigure[k-core]{ \includegraphics[width=.3\linewidth]{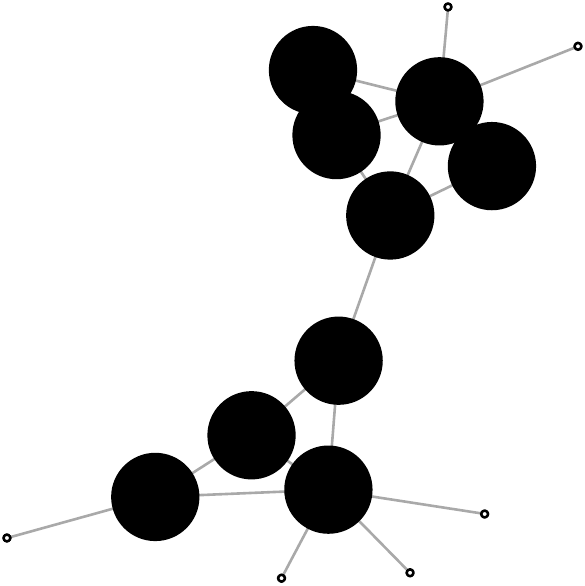}}
\subfigure[Closeness centrality]{ \includegraphics[width=.3\linewidth]{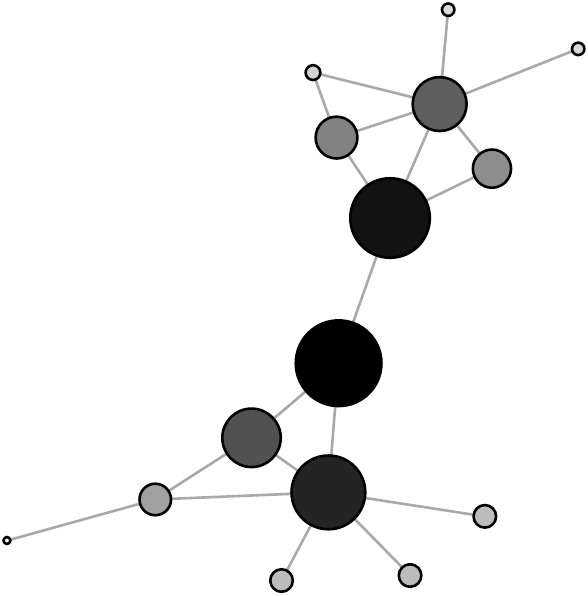}}
\subfigure[Betweeness centrality]{ \includegraphics[width=.3\linewidth]{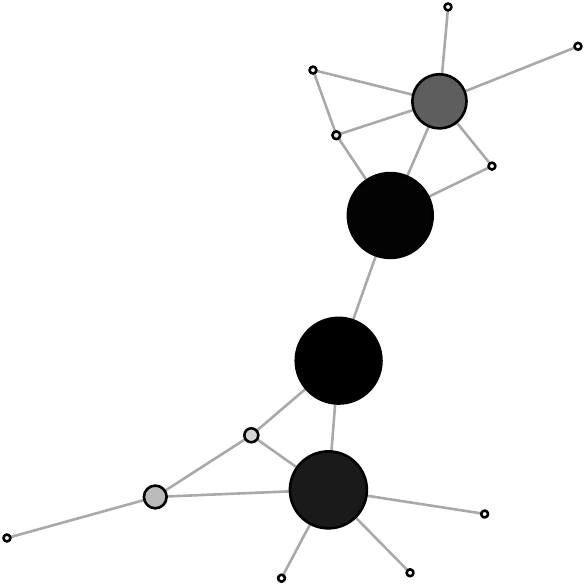}}
\subfigure[Eigenvector centrality]{ \includegraphics[width=.3\linewidth]{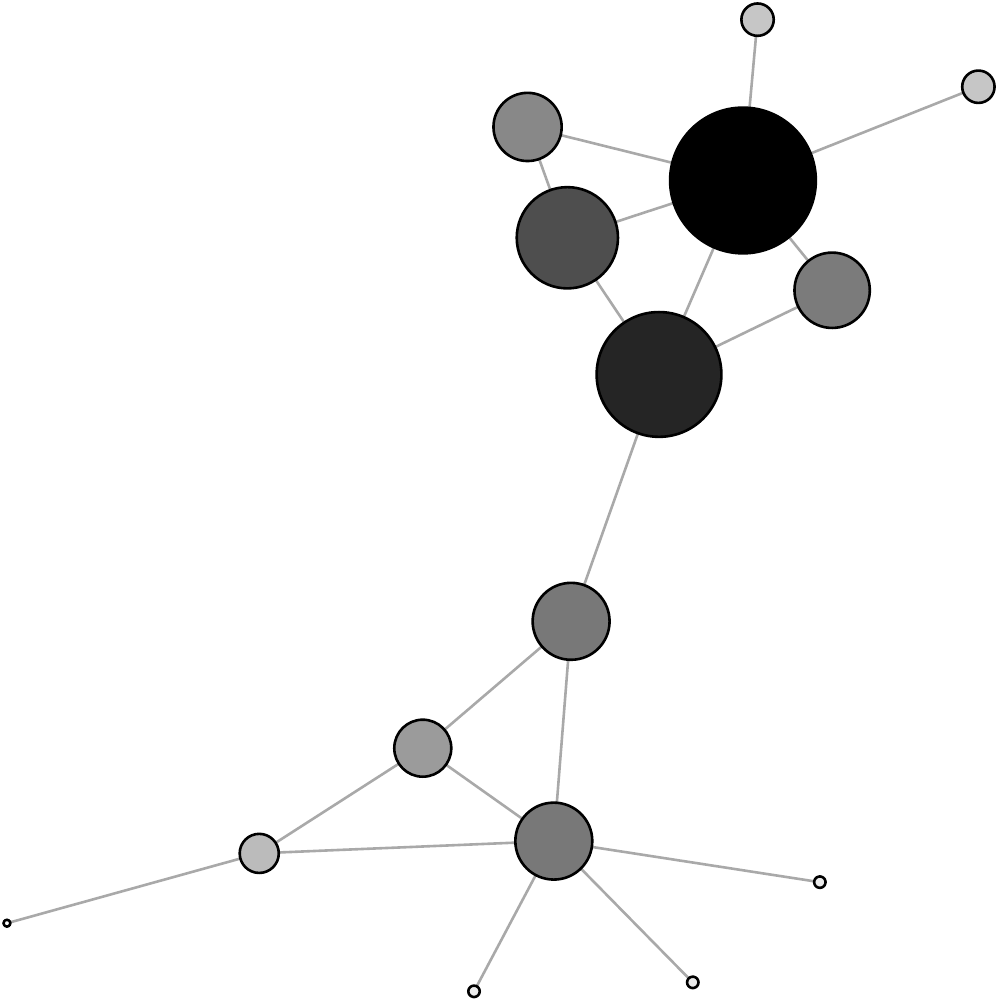}}
\subfigure[Acessibility]{ \includegraphics[width=.3\linewidth]{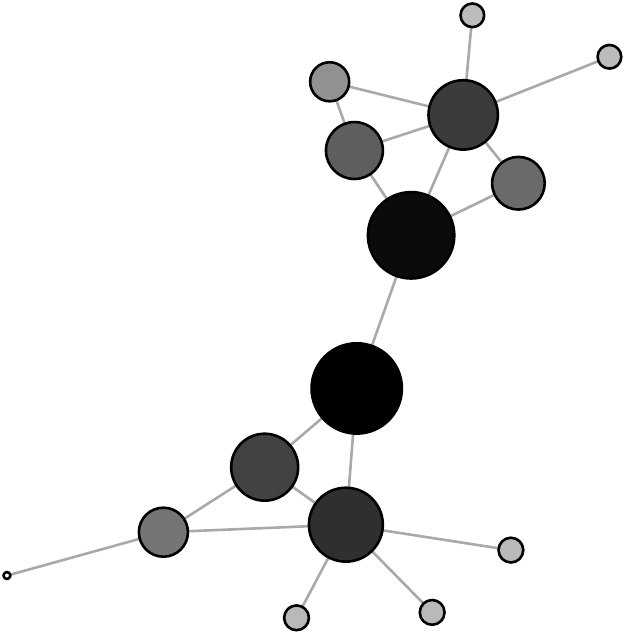}}
\end{center}
\caption{Comparison of centrality measures: (a) degree centrality, (b) k-core, (c) closeness centrality, (d) betweenness centrality, (e) eigenvector centrality and (f) accessibility. The size and darkness of each node are proportional to its centrality measure.}
\label{fig:centrality}
\end{figure}

The most simple centrality measure is the degree centrality, which is defined by the number of connections attached to each node. In terms of the adjacency matrix $A$, which is a mathematical representation of network structure ($A_{ij} = 1$ if there is a connection between nodes $i$ and $j$ or $A_{ij} = $ otherwise), we can calculate the degree centrality of node $i$ by the sum of the elements of row $i$ of $A$, i.e.,
\begin{equation}
k_i = \sum_{j=1}^N A_{ij},
\end{equation}
where $N$ is the number of nodes in the network. Figure~\ref{fig:degree-ex} shows an example of a network and the degree of each node. Although the degree definition is intuitive, since it is expected that highly connected nodes are at the center of the network, it has some drawbacks. For instance, as we can see in Figure~\ref{fig:centrality}(a), the nodes with the highest degree (in black) are at the periphery of the network and, therefore, are not central. Thus, the degree centrality can be considered as a local centrality measure, as a hub (densely connected node) may not be central.

A more general definition of centrality in terms of the number of connections is given by the $k$-core measure~\cite{Dorogovtsev06}. The $k$-core is a subgraph in which all vertices have a degree at least $k$. This centrality measure is obtained by the k-shell decomposition, which partitions the network by iteratively removing all nodes whose degree is smaller than $k$. After removing these nodes, the network is re-analyzed to verify whether there are nodes with less than $k$ connections. If such nodes are present, then they are also removed. The process is repeated until the minimum degree in the network is $k$. The resulting subgraph is called the $k$-core of the network. A node $i$ has coreness $kc(i) = k$ whether it belongs to the $k$-core, but it is not in the $(k+1)$-core. According to this measure, the most central nodes have the highest values of coreness ($k$-core number). Notice that high-degree nodes localized in the periphery of networks should display small values of $k$-core number since they are not in the main connected component when low degree nodes are removed. Therefore, only hubs at the main core of networks present the highest values of $kc$. The limitation of this measure lies in the fact that many nodes may be assigned to the same $k$-core number as we can see in Figure~\ref{fig:centrality}(b).

Node centrality can also be defined in terms of the shortest paths. The distance between nodes $i$ and $j$ is given by the number of edges in the shortest path connecting them. A central node is close to all other nodes in the network in terms of this distance. This idea is enclosed in the closeness centrality measure, which is defined in terms of the average distance of each node to all others. Mathematically, the closeness centrality of $i$ is defined as
\begin{equation}
C_i = \frac{N}{\sum_{j=1, j \neq i}^N d_{ij}},
\end{equation}
where $d_{ij}$ is the length of the shortest path between $i$ and $j$, and $N$ is the number of nodes in the network. Closeness centrality has the advantage to be very intuitive and suitable to characterize a process in which the information travel through the shortest distances. However, similar to degree and k-core, closeness centrality also presents some limitations. It is based only on the shortest distances and, therefore, the range of variation is too narrow due to the small diameter of networks. Indeed, most complex networks present small average shortest path length, since the typical distance increases with the logarithm of the number of nodes. The typical distance in a random network scales as $d \approx \log(N)/ \log(\langle k \rangle)$, where $\langle k \rangle = \sum_i \sum_j A_{ij}/N$ is the average degree. Thus, the ratio between the largest and minimal distances is of order $\log N$, since the minimal distance is equal to one. In most real-world networks, this ratio is about six or less. Thus, we can have several nodes with the same level of centrality, although they may present different roles on information spreading. This measure is more suitable when we have spatial networks, whose distance between nodes is higher than in random networks with the same number of nodes and connections.

If we consider the flow of particles on a network, then we can define centrality in terms of the load. It is natural to think that the most central node receives the largest number of particles in a defined time interval. Assuming that these particles move following the shortest distances, the load in a node $i$ is given by the total number of shortest paths passing through $i$. However, since we can have more than one shortest path between a pair of nodes $a$ and $b$, it is more suitable to define the load in node $i$ as the fraction of shortest paths connecting each pair of nodes $(a,b), a, b = 1, \ldots, N$, that includes $i$. Thus, mathematically~\cite{Freeman77},
\begin{equation}
B_i = \sum_{(a,b)} \frac{\eta (a,i,b)}{\eta (a,b)},
\label{betweenness}
\end{equation}
where $\eta (a,i,b)$ is the number of shortest paths connecting vertices $a$ and $b$ that pass through vertex $i$ and $\eta (a,b)$ is the total number of shortest paths between $a$ and $b$. The sum is over all pairs $(a,b)$ of distinct vertices. In this case, a central node is crossed by many paths and yields the highest value of $B$. 

This definition of betweenness centrality considers only the shortest distances and, therefore, is not general, since information can travel long distances in a network, such as occurs on the Internet, where routers adjust the route of IP packets according to the traffic flow. Moreover, the calculus of betweenness centrality is computationally expensive, requiring $O(N^3)$ time and $O(N^2)$ space, where $N$ is the number of nodes in the network. Even the solution proposed by Brandes~\cite{Brandes01} to calculate exact betweenness centrality, which runs in $O(NM)$, where $M$ is the number of edges in the network, is computationally expensive for large graphs. To overcome these limitations, calculus based on random walks can be considered~\cite{Newman05}. The betweenness centrality based on random walks is given by the expected number of visits to each node $i$ during a random walk,
\begin{equation}
B_i = \sum_{a=1}^N \sum_{b=1}^N w(a,i,b)
\end{equation} 
where $w(a,i,b)$ is the number of times that $i$ is visited when a random walk of length $n$ is performed from nodes $a$ to $b$. The number of steps to perform the random walk $n$ is a parameter of this method and the solution is, therefore, approximated. 

\begin{figure}[!t]
\begin{center}
\includegraphics[width=.9\linewidth]{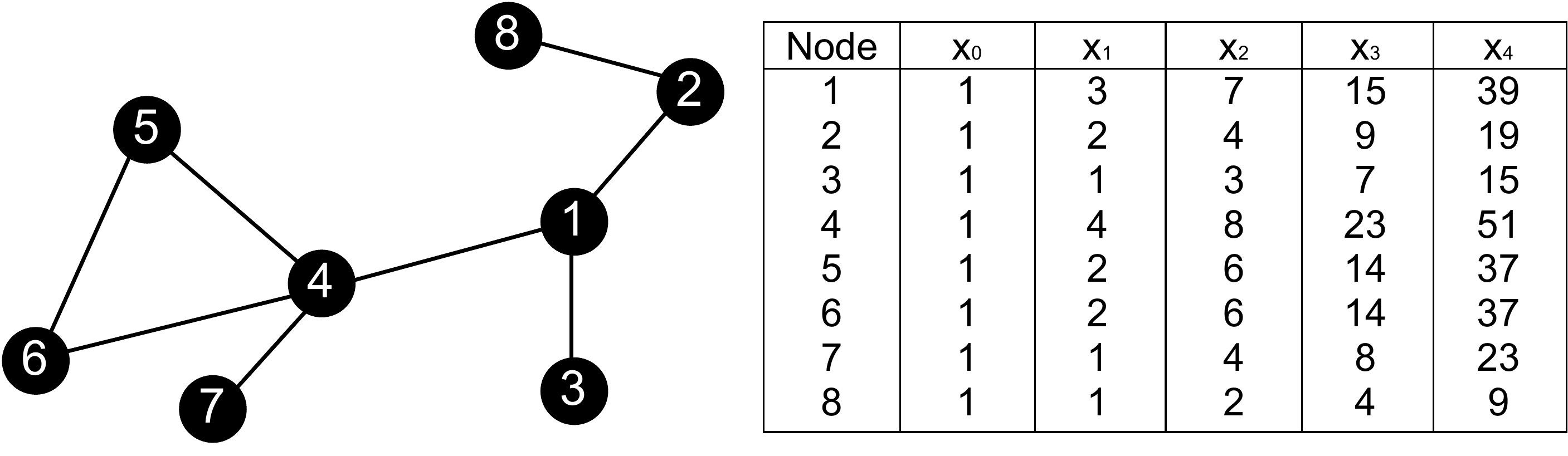}
\end{center}
\caption{Example of eigenvector centrality calculation. The values of $x$ are obtained by the power method (Equation~\ref{Eq:power}). We do not include normalization to make the example more didactic. Notice that $x_1$ stores the node degree. Although nodes 7 and 8 have the same degree, their eigenvector centrality indices are very different.}
\label{fig:eigen}
\end{figure}

We can also define centrality in terms of the neighborhood of each node. We can assume that a node is important if it is linked to other important nodes. For instance, a person in a social network, like Twitter, can be considered important if other important people follow him/her. Thus, if this ``importance'' is quantified by vector $x$, then we can define the importance of $i$ by
\begin{equation}
x_i = \frac{1}{\lambda} \sum_{k=1}^N A_{k,i} \, x_k,
\end{equation}
where $\lambda \neq 0$ is a constant. In matrix form we have: 
\begin{equation}
\lambda x = x A.
\end{equation}
That is, the ``importance'' of node $i$ is defined by the left-hand eigenvector of the adjacency matrix $A$ associated with the eigenvalue $\lambda$. The entries of $x$ are called eigenvector centrality. The eigenvalue $\lambda$ is chosen as the largest eigenvalue in the absolute value of matrix $A$, since due to the Perron-Frobenius theorem if matrix $A$ is irreducible then $x$ is both unique and positive. Since $\lambda$ is leading eigenvector of $A$, the calculation of $x$ can be performed by the power method,
\begin{equation}
x^{(k)} = x^{(k-1)} A,
\end{equation}\label{Eq:power}
where $x^{(0)}$ is an arbitrary vector with positive entries. If we consider $x^{(0)} = \{1,1,\ldots,1\}$, then $x^{(1)}$ is the degree and $x^{(n)}$, $n \leq 1$, is the number of walks of length $n$ arriving in each node (see Figure~\ref{fig:eigen}). Indeed, the number of walks between pairs of nodes is calculated from the power of the adjacency matrix. More specifically, the number of walks of length $n$ between every pair of nodes $i$ and $j$ is given by the entries of the matrix $A^n$. For instance, in Figure~\ref{fig:degree-ex} we can see the matrices $A^2, A^3$ and $A^4$. Therefore, the eigenvector centrality is proportional to the number of visits in each node through a random walk of infinite length. If a node is at the center of the network, then it is more accessed than the other nodes. Figure~\ref{fig:eigen} illustrates the calculation of the eigenvector centrality through the power method. Notice that nodes with the same degree, like nodes 7, 3 and 8, have different values of eigenvector centrality. Nodes at the center of the network present the highest values of eigenvector centrality.

Eigenvector centrality also presents some limitations. Depending on the network structure, most of the weights of the eigenvector concentrate in a few nodes, like hubs. In this case, most of the nodes will present centrality close to zero and, therefore, the importance of nodes is not well quantified. For instance, in a random network with only one hub, Martin et al.~\cite{Martin014} verified that in the localize regime, i.e., for $k > \langle k \rangle(\langle k \rangle + 1)$, where $k$ is the degree of the hub, the hub vertex and its neighbors present the highest values of eigenvector centrality, whereas the centrality of the remaining nodes is $O(1/N)$, vanishing in the limit of large networks. After this value, the inverse participation ratio defined as
\begin{equation}
IPR = \frac{\sum_{i=1}^N x_i^4}{(\sum_{i=1}^N x_i^2)^2},
\end{equation}
where $x_i$ is the eigenvector centrality of node $i$, presents a phase transition, indicating the presence of localization. The value of $IPR$ is close to $1$ if $x$ is localized.

\begin{figure}[!t]
\begin{center}
\includegraphics[width=.6\linewidth]{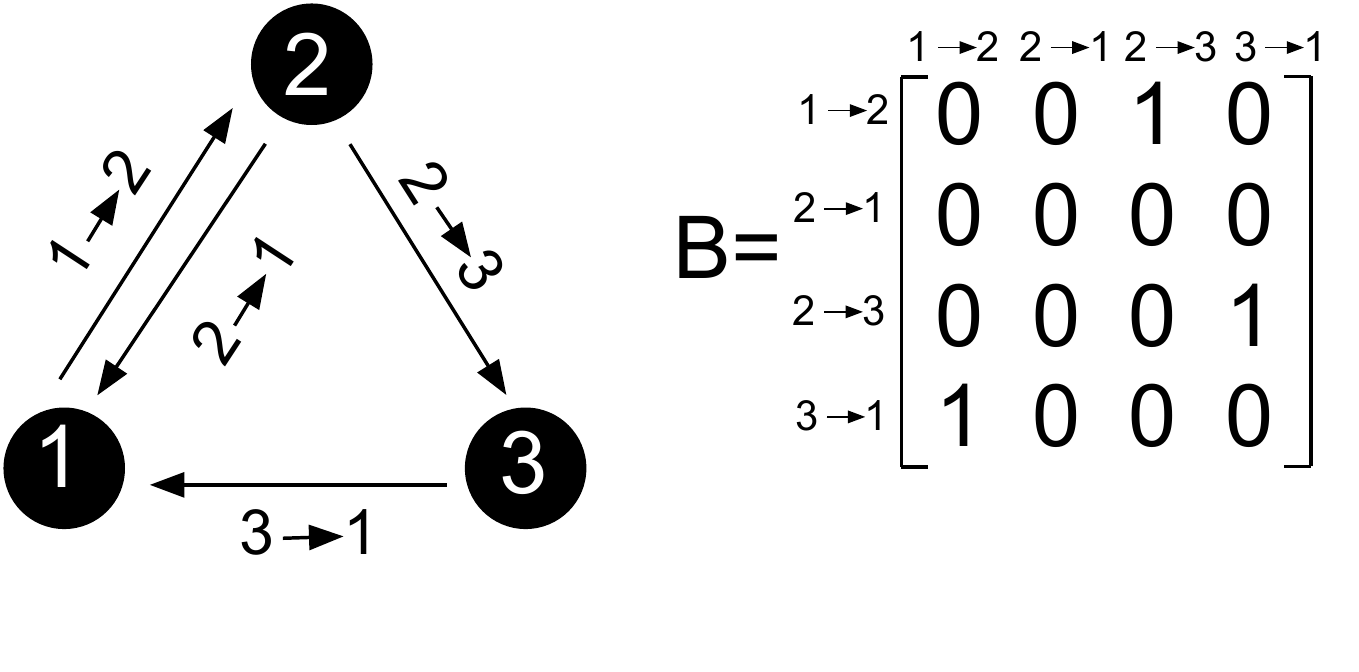}
\end{center}
\caption{Illustration of the non-backtracking matrix. If there is a link $i\rightarrow j$ and $j \rightarrow k$, then $B_{i\rightarrow j, j \rightarrow k} = 1$. }
\label{fig:backtracking}
\end{figure}

The limitation of the eigenvector centrality can be overcome by considering the Hashimoto non-backtracking matrix $B$. $B$ is a $2M\times 2M$ non-symmetric matrix with one row and one column for each directed edge $(i,j)$. For undirected networks, each undirected link between $i$ and $j$ is replaced with a pair of directed links $(i,j)$ and $(j,i)$. This matrix is defined for undirected networks by replacing each undirected link between $i$ and $j$ with a pair of directed links $(i,j)$ and $(j,i)$ and make $B$ a $2M\times 2M$ non-symmetric matrix with one row and one column for each directed edge $(i,j)$. The elements of $B$ are defined by
\begin{equation}
B_{u\rightarrow v,w \rightarrow x} = 
\left\{ 
\begin{array}{l}
1  \mbox{ if } v = w \mbox{ and } u \neq x\\ 
0 \mbox{ otherwise}.
\end{array}
\right. 
\end{equation}
The entry $v_{i \rightarrow j}$ of the main eigenvector $v$ of $B$ gives the centrality of vertex $i$ disregarding the contribution of node $j$. Thus, the centrality of node $j$ is given by the sum of these contributions for all neighbors of $j$, i.e., $x_j = \sum_{i=1}^N A_{ij} v_{i \rightarrow j}$. Figure~\ref{fig:backtracking} illustrates the calculation of the non-backtracking matrix. In practice, the calculation of the centrality based on the non-backtracking matrix can be performed by considering the so-called Ihara (or Ihara--Bass) matrix~\cite{Martin014}
\begin{equation}
B = \left(
\begin{array}{cc}
0 & D-I\\ 
-I & A
\end{array}
\right),
\end{equation}
where $I$ is the $N \times N$ identity matrix, $D$ is a diagonal matrix whose element is equal to the node's degree $Dii = \sum_{j=1}^N A_{ij}$. The vector $x$ of centralities is equal to the first $N$ entries of the leading eigenvector of the $2N \times 2N$ matrix.

The name non-backtracking is due to non-backtracking walks, which is a walk in which it is allowed to return to a vertex visited previously only after at least two other vertices have been visited. For instance, a sequence like $i\rightarrow j \rightarrow i$ is not allowed. The entries $(B_{uv})^n$ yield the number of non-backtracking walks of length $n$ between the edges $u$ and $v$ in the original network, in the same way as powers of the adjacency matrix count the number of paths between pairs of nodes (see Figure~\ref{fig:degree-ex}). These properties of $B$ make it useful for community detection in networks by using spectral clustering~\cite{Krzakala013}. In this case, when the network presents homophily, i.e., nodes in the same community are more likely connected than between communities, the eigenvalue associated with this eigenvector is separated from the bulk of the eigenvalue spectrum and communities can be easily obtained. However, when homophily is not well defined, the relevant eigenvalue dissolves into the bulk and the community identification is impossible. Thus, we have a detectability threshold for community detection in which no algorithm can label the vertices better than chance. Nevertheless, by using the non-backtracking matrix, Krzakala et al.~\cite{Krzakala013} show that it is possible to detect communities above this threshold. 
They verified that spectral algorithms based on the non-backtracking matrix perform optimally for some popular generative models, including the stochastic block model. A complete discussion about community identification methods, including those based on matrix $B$ is presented in~\cite{Fortunato016}.


Although the non-backtracking matrix overcomes the localization problem, it presents some drawbacks. Since the size of $B$ is $2M \times 2M$, the computation of its eigenvalues and eigenvectors is computationally expensive. Thus, the calculation of $B$ is intractable for networks with millions of vertices. Moreover, non-backtracking walks must contain cycles and, therefore, subgraphs representing trees do not contain non-backtracking walks. These trees have no effect on the spectrum and can be removed from the network. However, this removal represents a loss of information about the network structure, as these subgraphs may be important for network characterization, like in community identification~\cite{Fortunato016}, and dynamical processes analysis, as in the case of synchronization~\cite{Schultz016}. An alternative version of the non-backtracking matrix, called flow matrix, was proposed by Newman~\cite{Newman013} to overcome these limitations. However, the manipulation of this matrix is also expensive, since it has the same dimension as the non-backtracking matrix.

Similar to the eigenvector centrality, PageRank is another measure based on random walks~\cite{Brin98}. In this case, the basic idea is to transform the adjacency matrix such that its elements represent the probability transition between a pair of nodes. The network can represent a Markov chain in which each node is a state. Since this chain can have absorbing or periodic states, the PageRank algorithm transforms it to become ergodic, such that each node can be reached from every other node by following random walks. This transformation is particularly important in directed networks, since the network may present absorbing nodes, where a random walker can be trapped. As we verified for the eigenvector centrality, the spectra of adjacency matrix yield the expected number of visits in each node by following a random walk. In the case of PageRank, we have a similar idea. Thus, the PageRank is calculated by the power method,
\begin{equation}
\pi^T = \pi^T G,
\end{equation}
where $G$ is the Google matrix, 
\begin{equation}
G = \kappa \left(P + \frac{ae^T}{N} \right) + \frac{(1-\kappa)}{N} e e^T.
\end{equation}
The element $a$ is the binary vector called dangling node vector ($a_i$ is equal to one if $i$ is a dangling node and $0$ otherwise), $e$ is a vector of ones of length $N$ and $P$ is the transition probability matrix of the respective network,
\begin{equation}\label{Eq:P}
P_{ij} = \frac{A_{ij}}{\sum_j A_{ij}}.
\end{equation}
The original version of the algorithm considers $\kappa = 0.85$~\cite{Brin98}. The PageRank of a node $i$, $\pi_i$, is given by the $i$-th entry of the dominant eigenvector $\pi$ of $G$, given that $\sum_i \pi_i = 1$. $\pi_i$ can be understood as the probability of arriving at the node $i$ after a large number of steps following a random walk navigation through the network.

One limitation of the PageRank measure lies in the definition of connections among all pair of nodes, which simulate the jumping after random walk navigation. This procedure selects all nodes with uniform probability, ignoring their importance. However, in real-world navigation, some web pages are more likely to be selected than other ones. For instance, pages related to news are more visited than personal  websites. 

Accessibility is another measure that considers random walks on networks. This measure is related to the diversity of access of individual nodes through random walks.  The accessibility of the node $i$ for a given distance $h$ is defined by the exponential of the Shannon entropy of the transition probability matrix (Equation~\ref{Eq:P})~\cite{Travencolo08}, i.e.,
\begin{equation}
\alpha_h(i) = \exp\left({-\sum_{j}P^{(h)}_{ij}\log P^{(h)}_{ij}}\right),
\label{Eq:acch}
\end{equation}
where $1 \leq \alpha_h(i) \leq N$. The elements of $P^{(h)}_{ij}$ yield the probability of going from $i$ to $j$ following a walk of length $h$. The maximum value of $\alpha_h$ corresponds to the case in which all nodes are reached with the same probability $1/N$. 

This version of accessibility depends on the parameter $h$. However, it can be generalized by considering the matrix exponential operation~\cite{Arruda014}, 
\begin{equation}\label{Eq:W}
\textbf{P}  = \frac{1}{e}\sum_{k=0}^{\infty} \frac{1}{k!} {P}^k = e^{P},
\end{equation} 
where the factor $1/e$ is necessary to guarantee that $\textbf{P}$ is a stochastic matrix~\cite{Arruda014}. Notice that the definition of $\textbf{P}$ penalizes long walks, i.e., the shortest walks receive more weight than the longest ones. The generalized random walk accessibility is defined as
\begin{equation}\label{Eq:acc}
\alpha(i) = \exp\left(-\sum_j \textbf{P}_{ij}\log \textbf{P}_{ij}\right),
\end{equation}
where $\textbf{P}_{ij}$ is calculated from Equation~\ref{Eq:W}.
This measure has been used to identify the border of networks~\cite{Travencolo09}. The limitation of accessibility lies in the penalization of longer walks. Moreover, it is assumed that the random walk selects nodes with the same probability in each step, which may not occur in real-world navigation.

In Figure~\ref{fig:centrality} we compare the centrality measures discussed in this section. As we can see, different centrality measures may not identify the same set of central nodes. Thus, the use of centrality metrics depends on the problem and the most suitable approach is to consider more than one centrality for network characterization. For instance, one can combine a set of measures and consider multivariate statistical methods for network characterization.

Although all centrality measures present some drawbacks, they are very useful in the analysis of dynamical processes on networks. Centrality is also a fundamental property of several complex systems, like brain networks, Internet, and social networks. As follows, we discuss the importance of centrality to understanding the structure and dynamics of complex systems and some application of centrality for characterization of the structure and function of complex systems.



\section{Centrality and dynamical processes in networks}

Centrality plays a fundamental role in the evolution of dynamical processes in networks, like epidemic spreading and synchronization. Due to the presence of hubs in scale-free networks, the epidemic threshold is close to zero in these structures~\cite{Pastor015}. Similarly, the critical coupling for the emergence of synchronization of Kuramoto oscillators also depends on the network heterogeneity, which decreases with the variation of the node degree~\cite{Rodrigues016}. In this section, we analyze these two dynamical processes in terms of centrality measures and point out the influence of central nodes on the evolution of the dynamical process.

\subsection{Identification of influential spreaders}

\begin{figure}[!h]
\begin{center}
\subfigure{ \includegraphics[width=.35\linewidth]{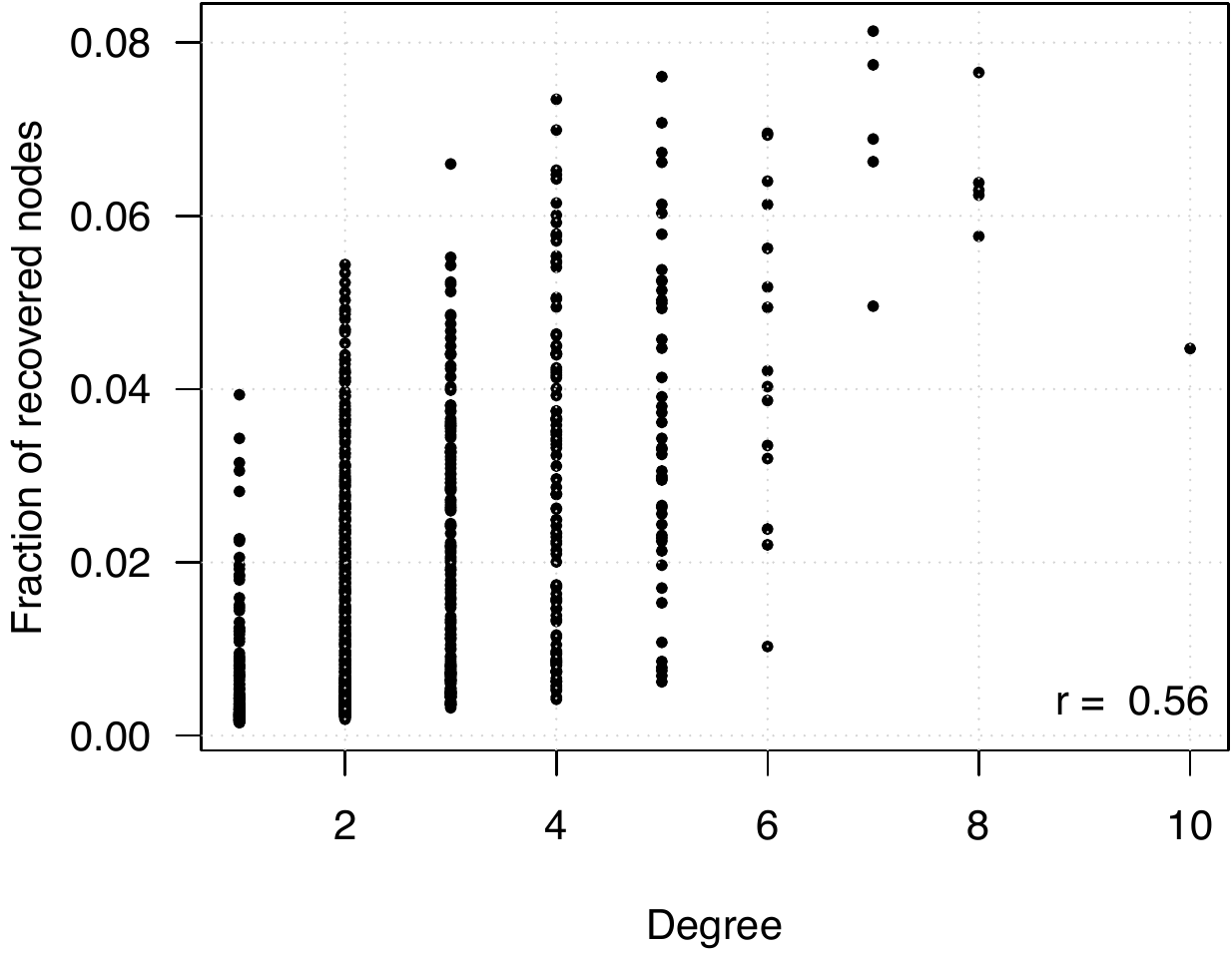}}
\subfigure{ \includegraphics[width=.35\linewidth]{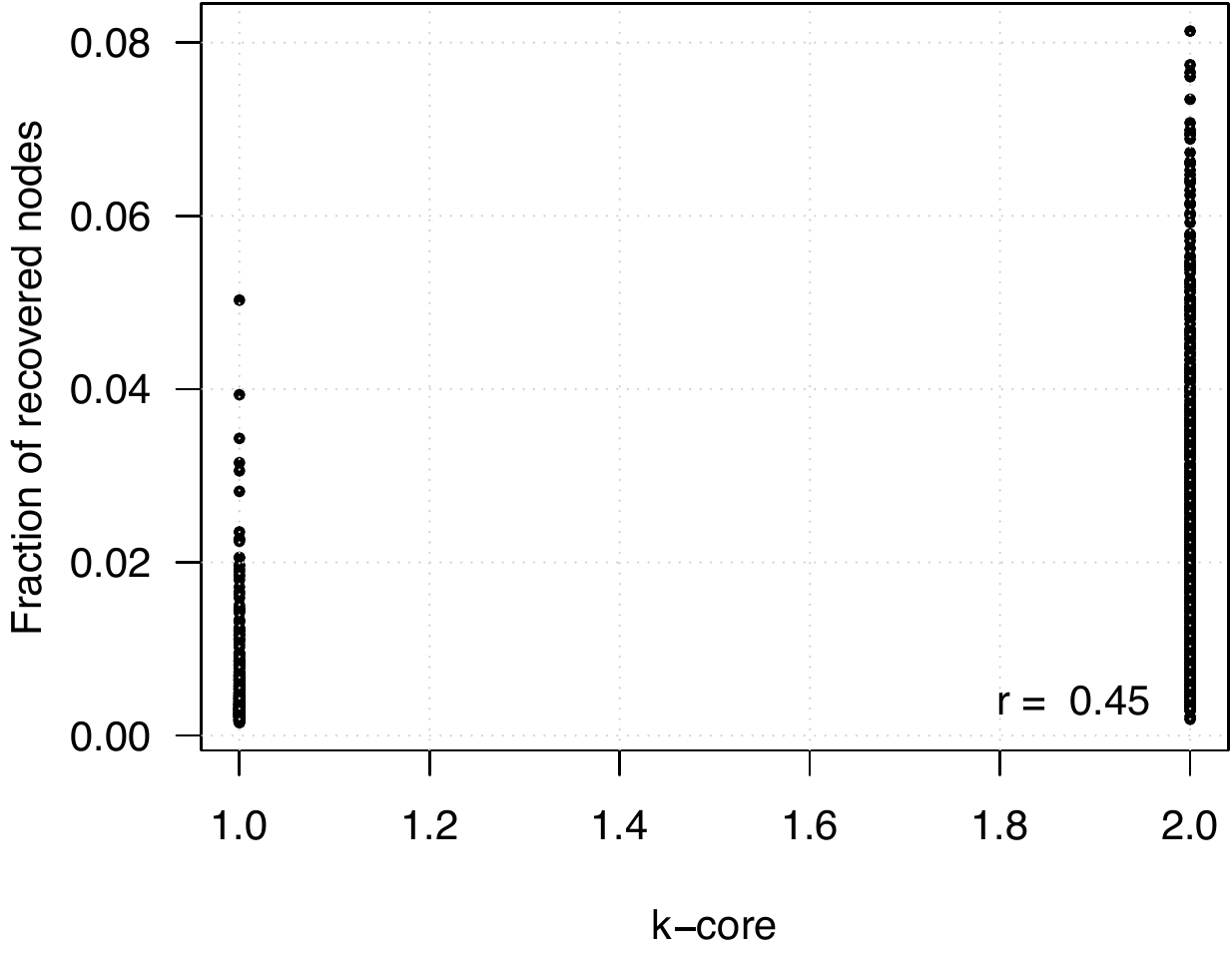}}
\subfigure{ \includegraphics[width=.35\linewidth]{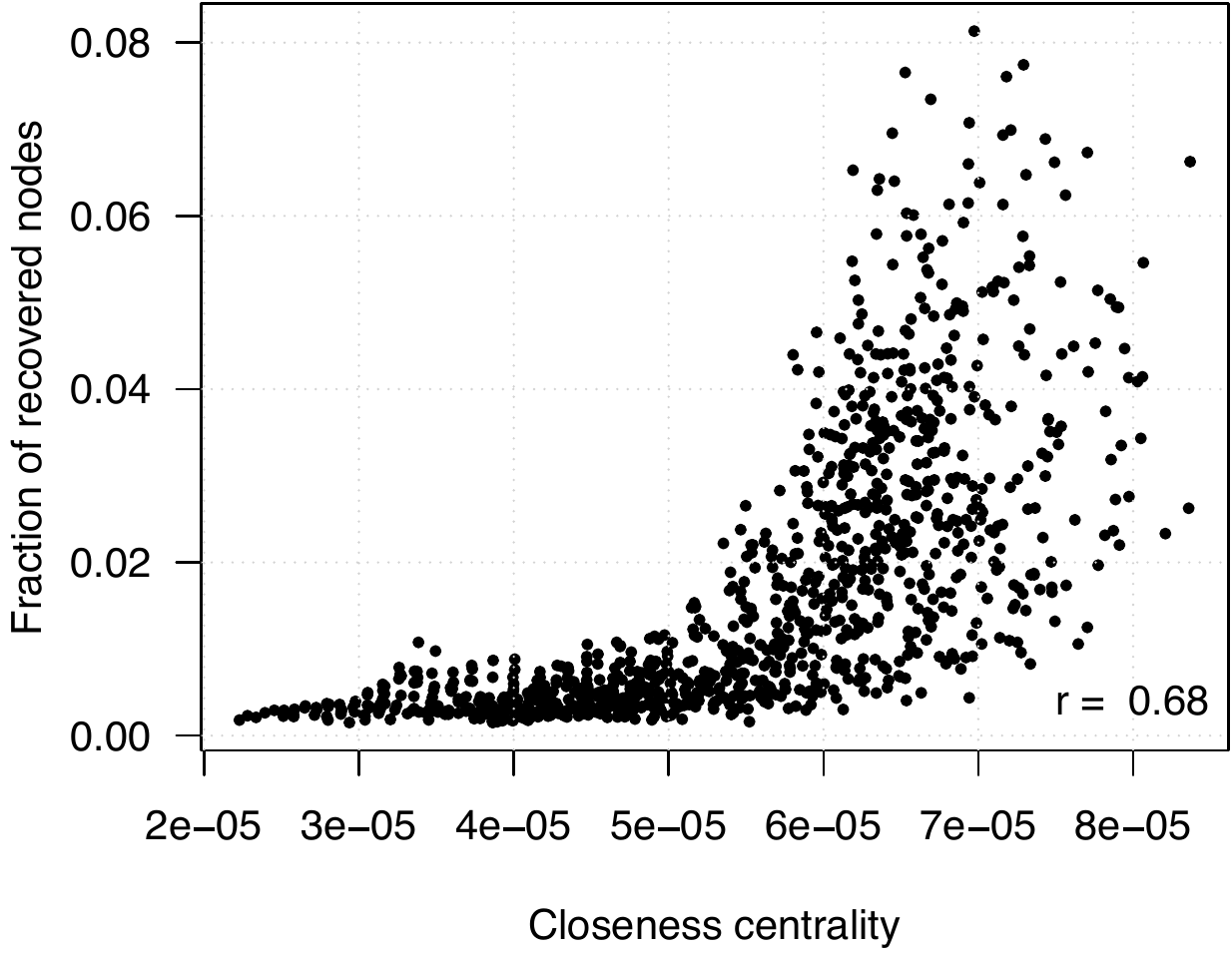}}
\subfigure{ \includegraphics[width=.35\linewidth]{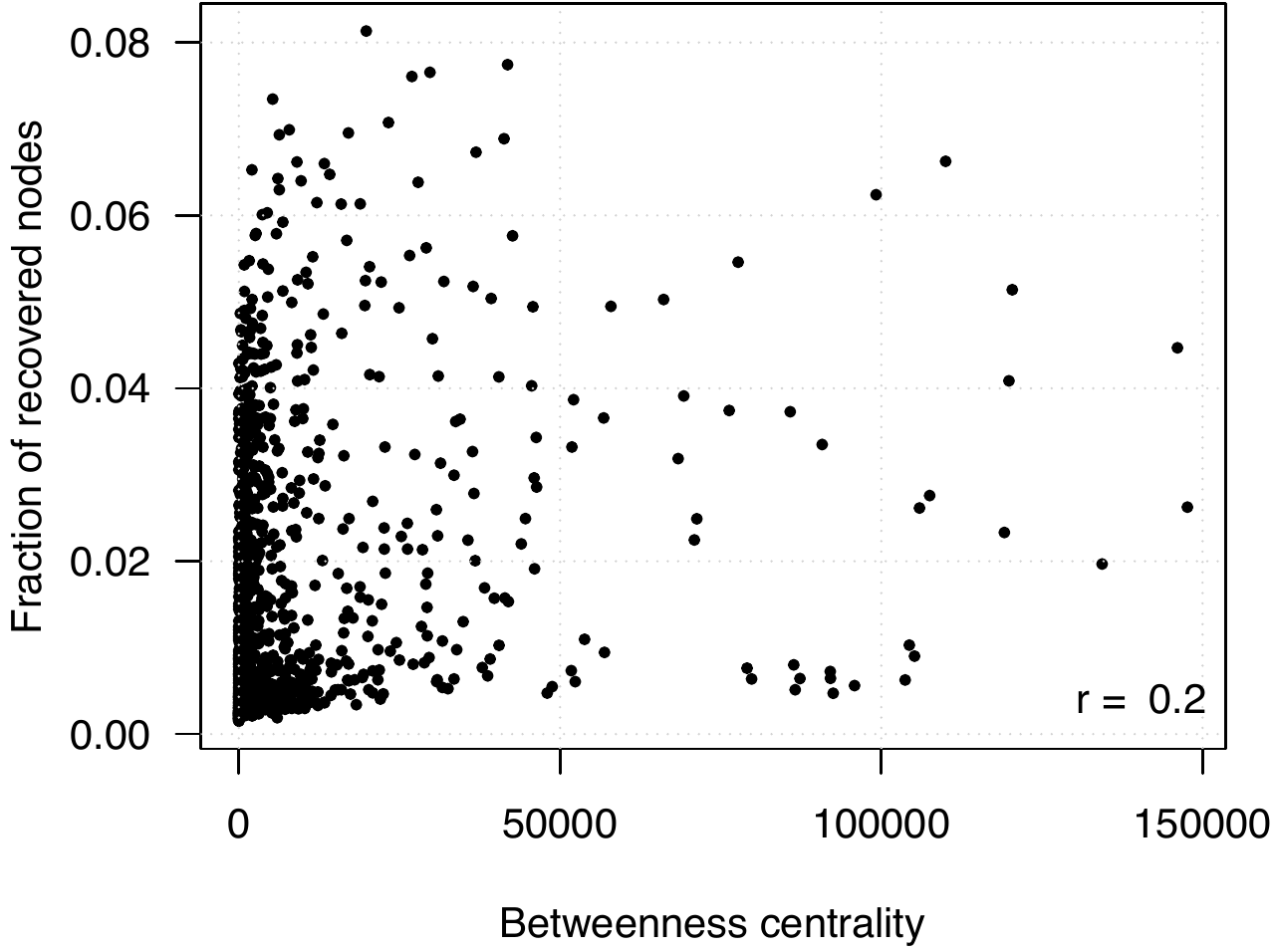}}
\subfigure{ \includegraphics[width=.35\linewidth]{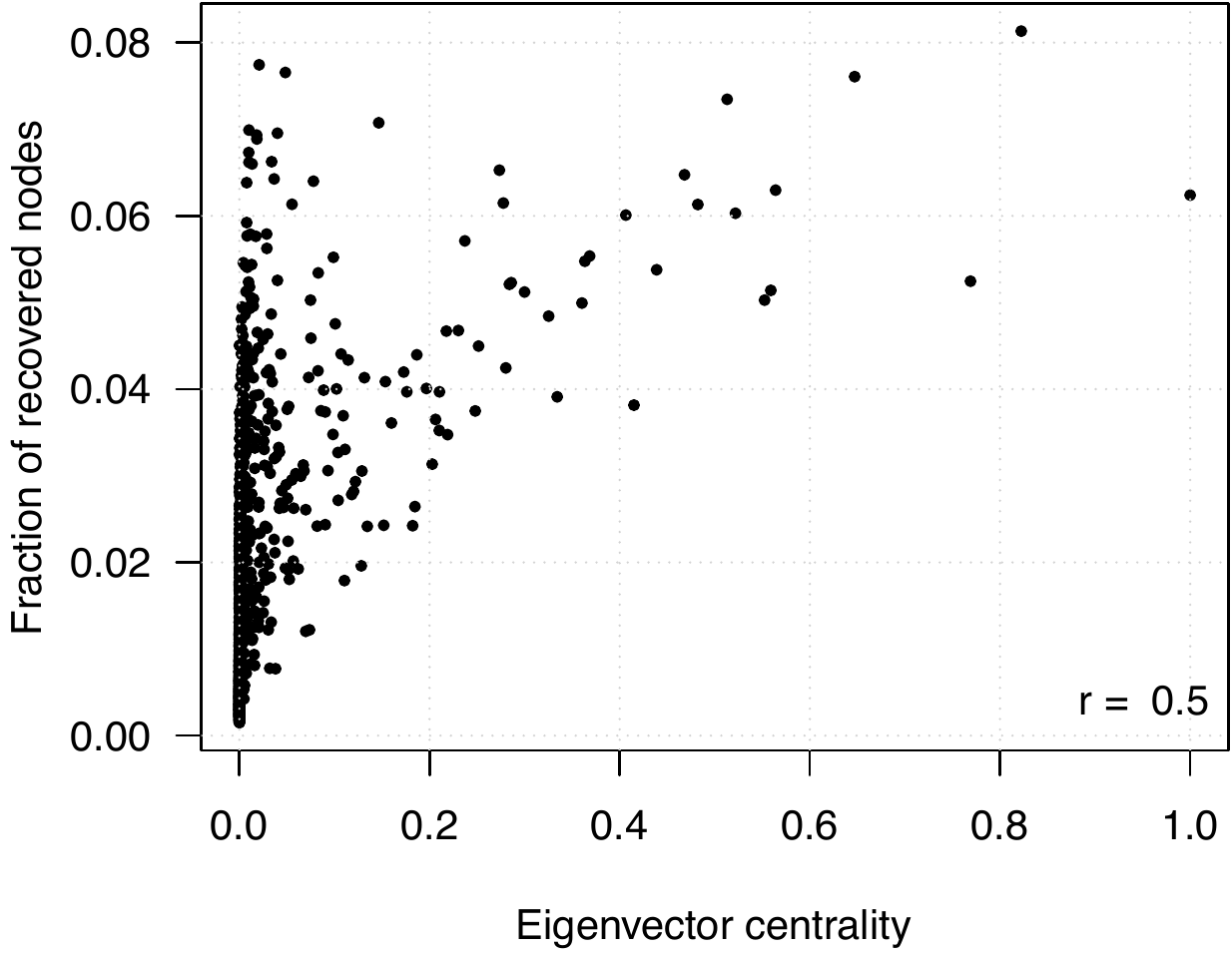}}
\subfigure{ \includegraphics[width=.35\linewidth]{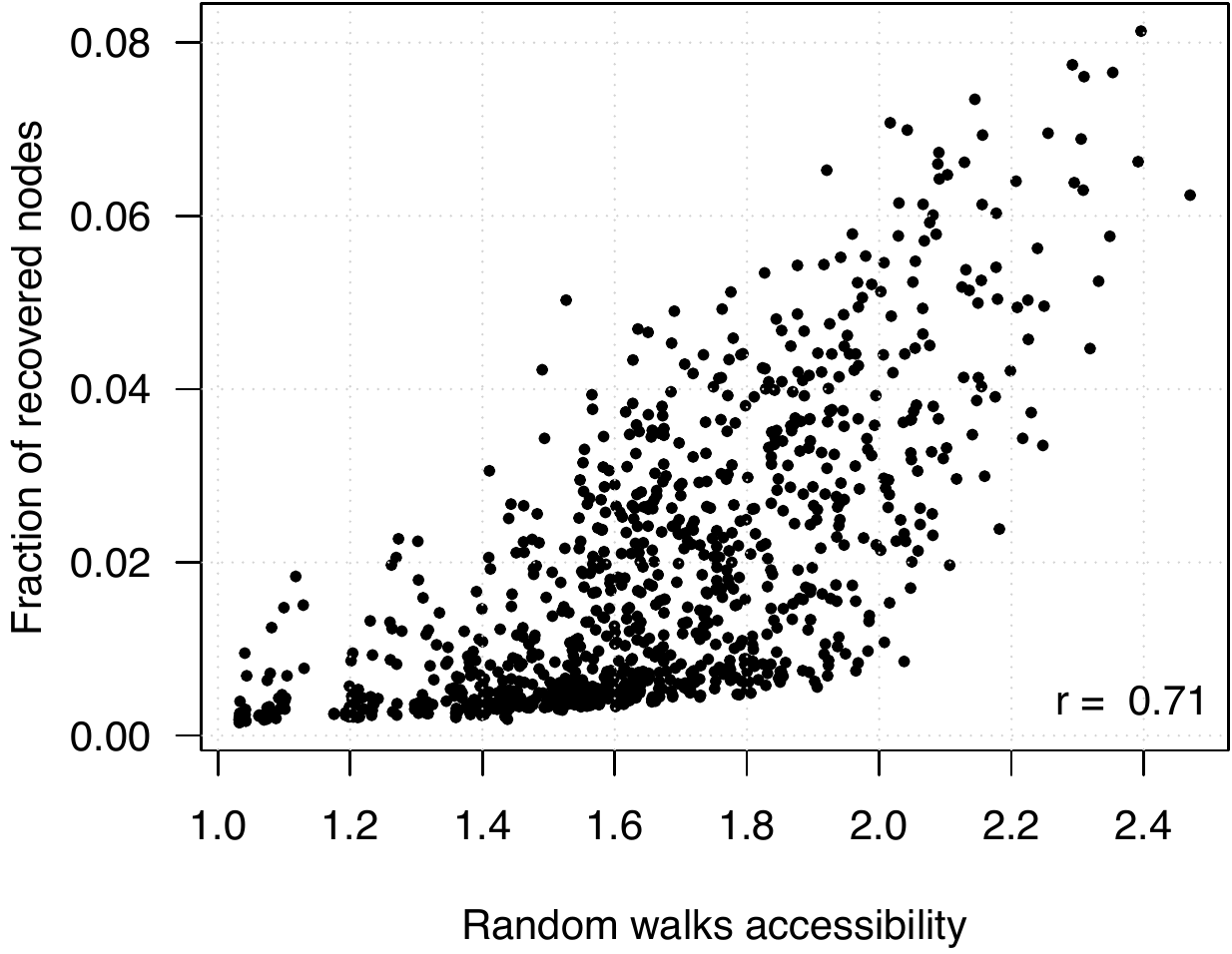}}
\end{center}
\caption{The relation between the outbreak size and centrality measures in the E-road network. The epidemic spreading is simulated through the SIR model with discrete time, recover probability $\mu = 1$, and spreading probability $\beta = 2\lambda_c$, where $\lambda_c = 1 / \lambda_{max}$ and $\lambda_{max}$ is the main eigenvalue of the adjacency matrix of the network. Each point represents a city with a given centrality and spreading capacity, given by the average outbreak size. This average is calculated from 50 simulations of the SIR model starting in each node. The network has $N = 1,174$ cities and $M = 1,417$ roads. The Pearson correlation coefficient between the percentage of recovered individuals and centrality measures is indicated inside the plots.  Only the network largest component is considered in our analysis.} \label{fig:road}
\end{figure}

Epidemic processes are ubiquitous in nature, society, and technology~\cite{Pastor015}. The theoretical study of infections disease aims to improve control and eradicate the pathogen agent from the population~\cite{Barrat08, Pastor015}.  Many mathematical models have been developed to understand the propagation of diseases in populations of humans and animals~\cite{Keeling08}. One particularly important model is the SIR (susceptible-infected-recovered) model, in which subjects can recover from the disease and acquire immunity. We consider this model in our analysis of centrality and epidemic outbreaks.
 
The importance of identification of the main propagators of disease lies in developing efficient methods for disease control. For instance, many recent works have analyzed the influence of centrality on epidemic spreading~\cite{Kitsak010, Arruda014, Radicchi016}. Here, we address this influence by the centrality measures described in Section~\ref{Sec:centrality}. We simulate the SIR epidemic spreading by following the reactive process~\cite{Gomez010}, in which at each time step, every infected node triesf to transmit the disease to all of its neighbors with a fixed probability $\beta$. After that, each of these infected nodes recovers with probability $\mu$. We consider $\mu = 1$ and $\beta = 2\lambda_c$, where $\lambda_c = \frac{1}{\Lambda_1}$ is the critical threshold of epidemic spreading and $\Lambda_1$ is the leading eigenvector of the adjacency matrix. This critical threshold depends on the network structure, as verified before in several works~\cite{Pastor015}. 

In Figure~\ref{fig:road}, we present the relation between the average outbreak size when the disease starts in each node and its respective centrality measure. For most of the measures, the most central nodes are the most influential spreaders. However, the generalized random walk accessibility is the best metric to predict the outbreak size in the E-road network, which is a road network located mostly in Europe whose nodes represent cities, and edges are the roads. K-core and degree poorly predict the most influential spreaders. This fact is mainly due to the spatial organization of the network, which does not present hubs and, therefore, does not have a scale-free organization. Betweenness centrality also cannot predict the most influential spreaders.

\begin{figure}[!t]
\begin{center}
\subfigure{ \includegraphics[width=.35\linewidth]{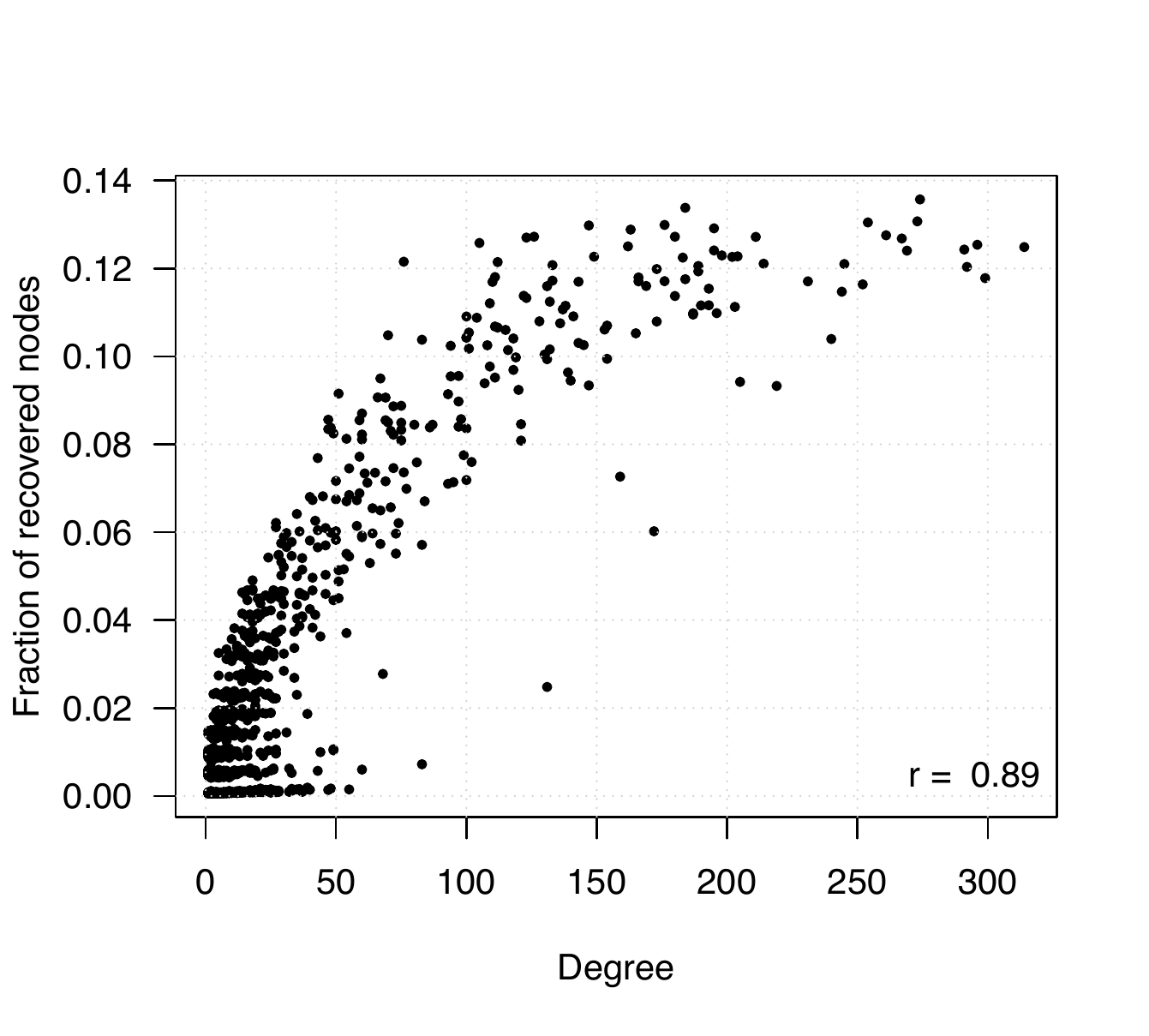}}
\subfigure{ \includegraphics[width=.35\linewidth]{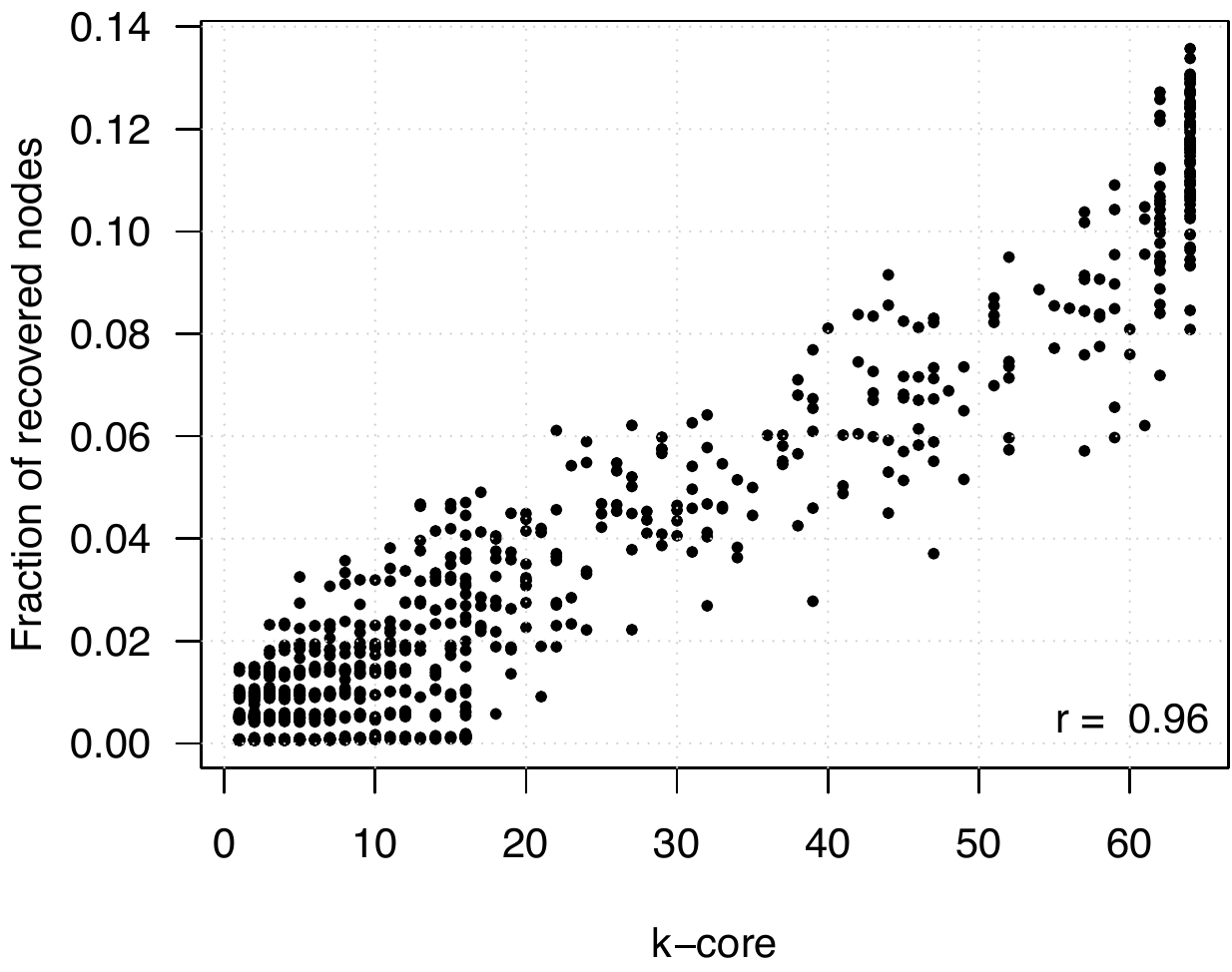}}
\subfigure{ \includegraphics[width=.35\linewidth]{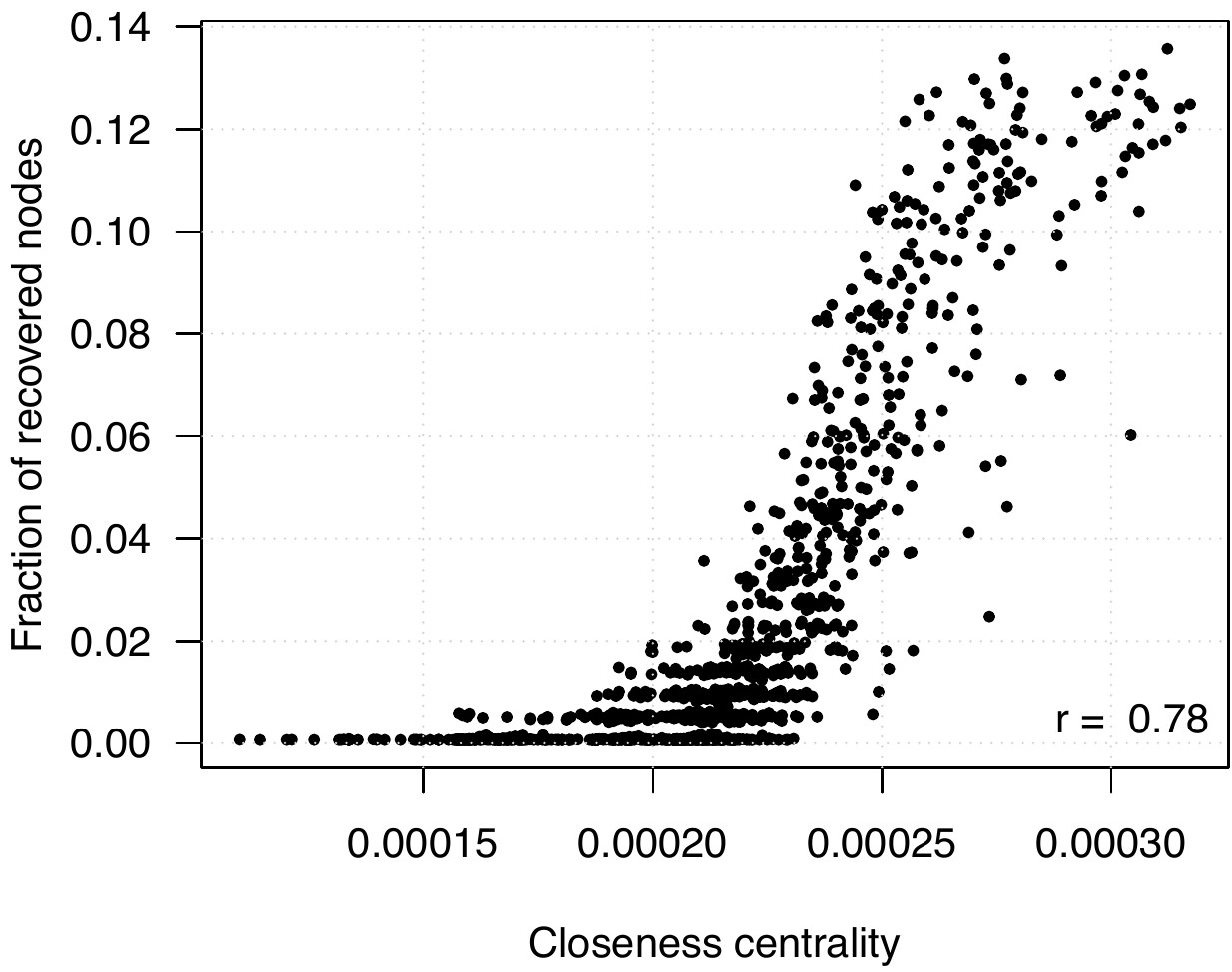}}
\subfigure{ \includegraphics[width=.35\linewidth]{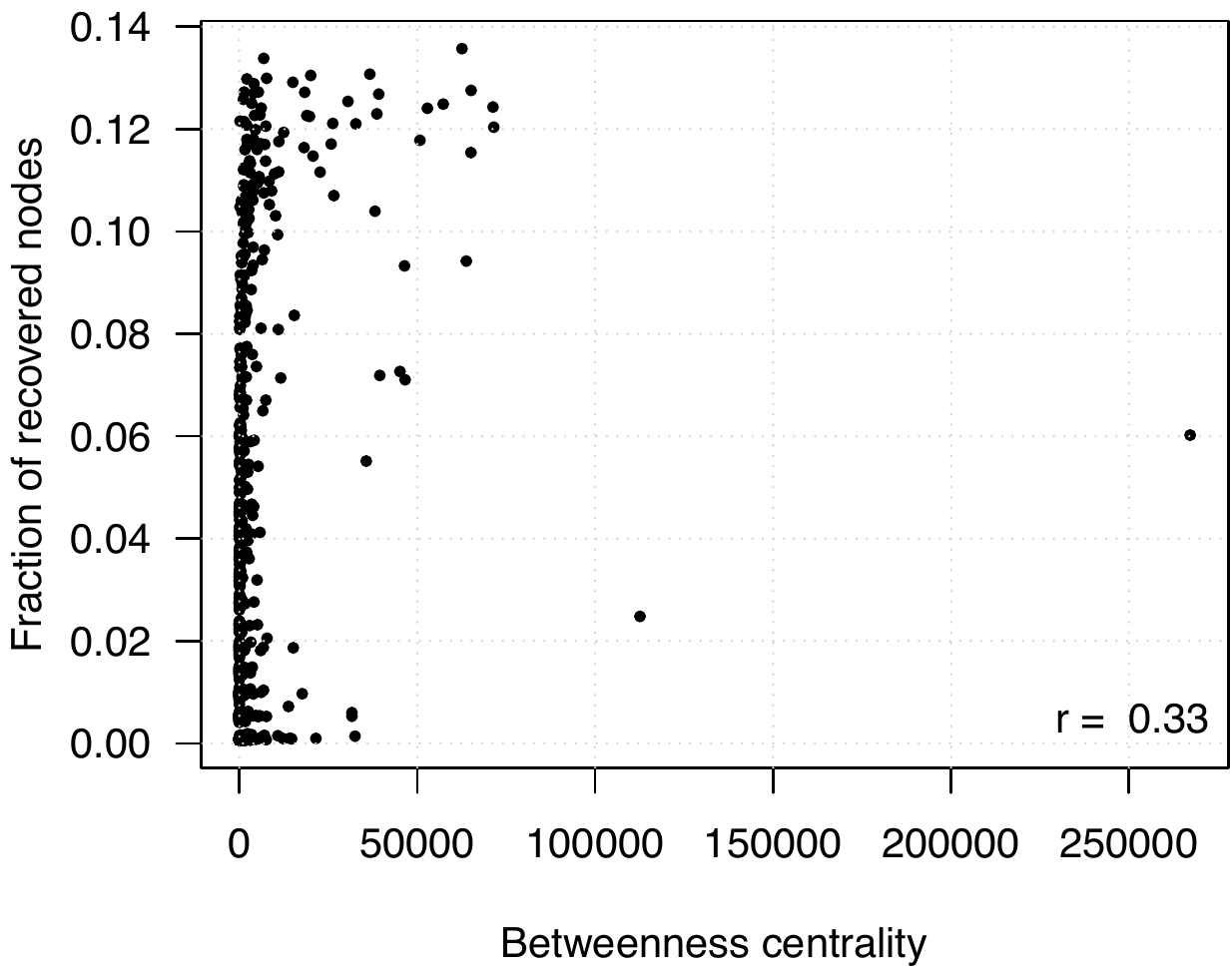}}
\subfigure{ \includegraphics[width=.35\linewidth]{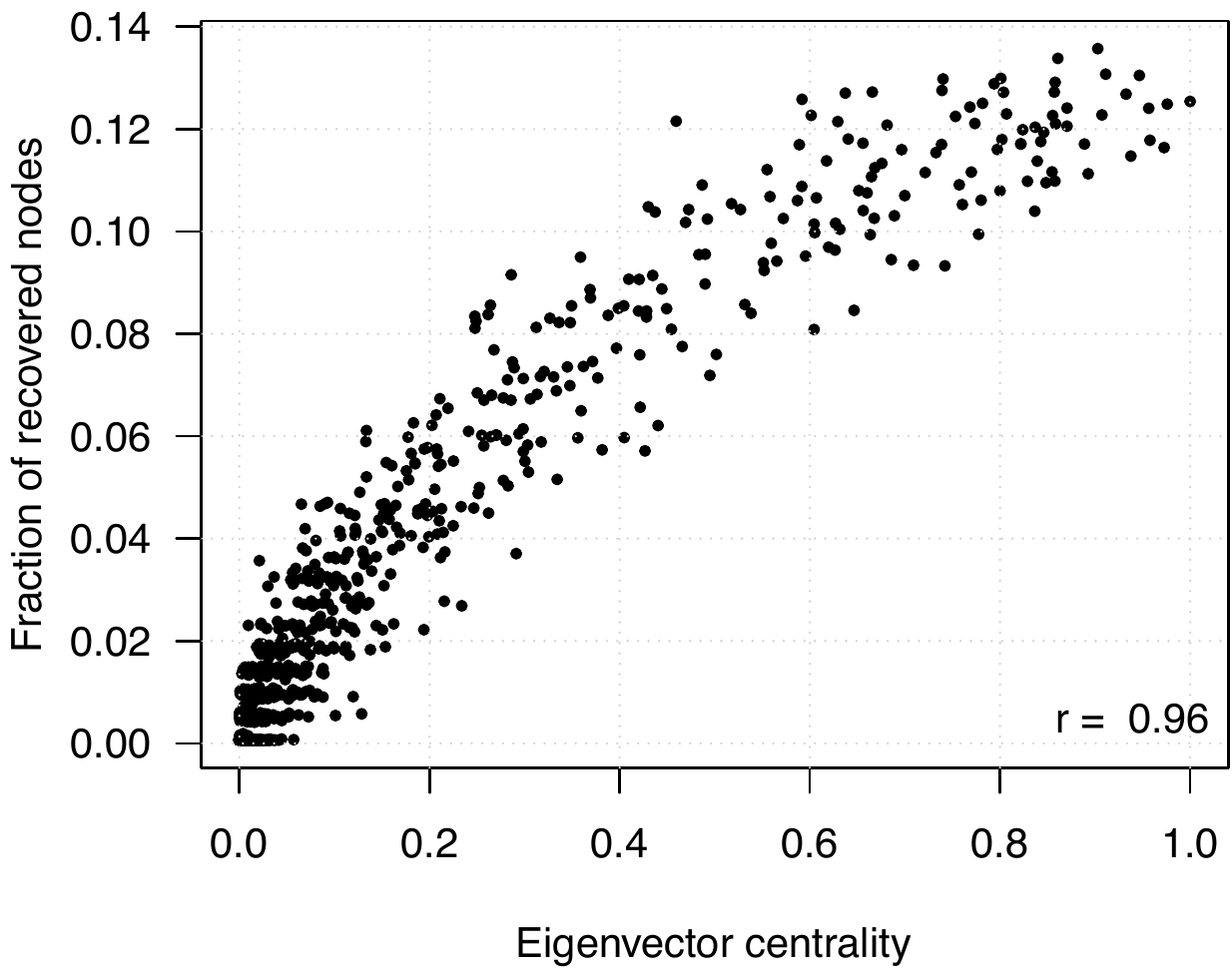}}
\subfigure{ \includegraphics[width=.35\linewidth]{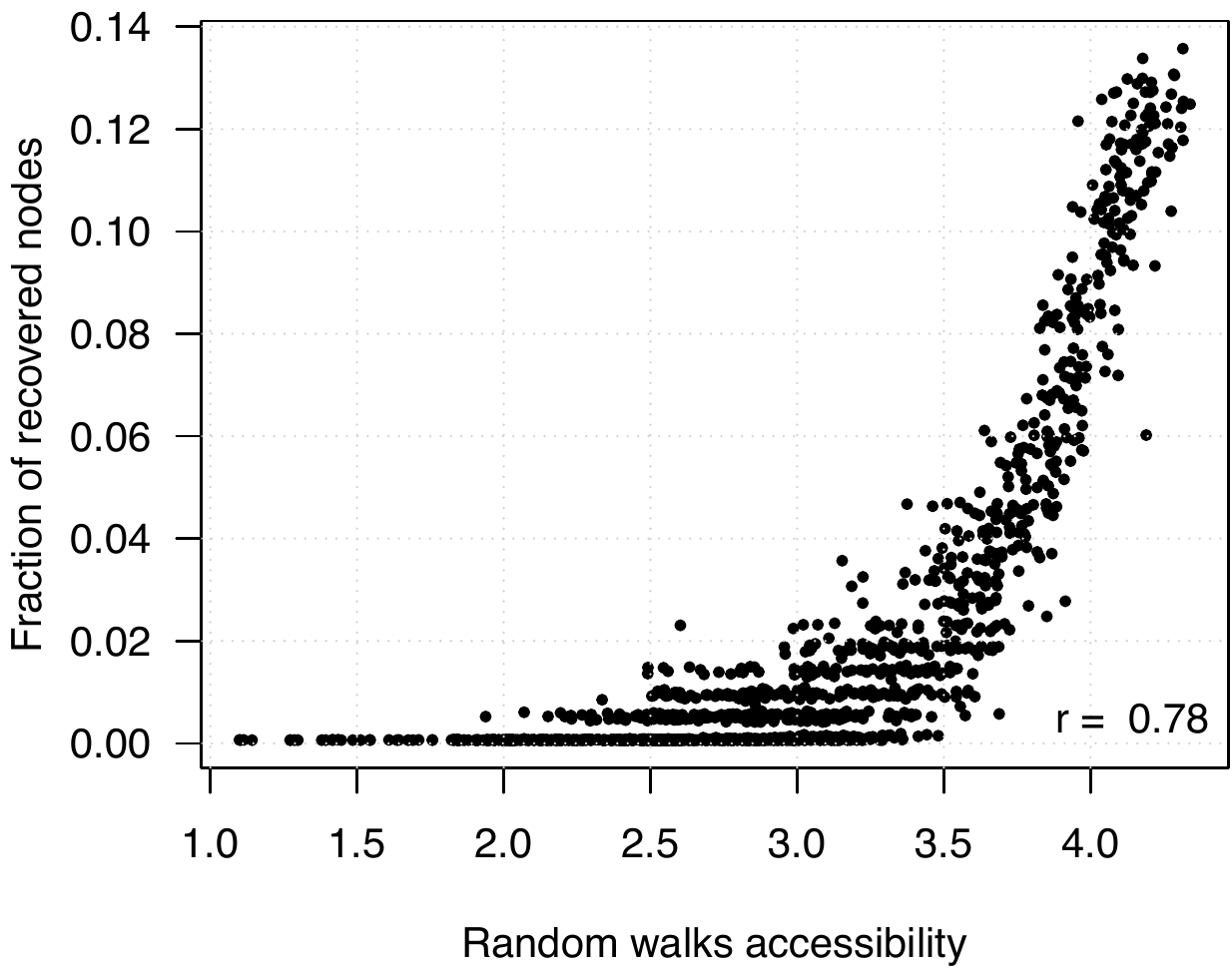}}
\end{center}
\caption{Outbreak size and centrality measures in the US airports network. The epidemic spreading is simulated through the SIR model with discrete time and contagion probability $\beta = 2\lambda_c$, where $\lambda_c = 1 / \lambda_{max}$ and $\lambda_{max}$ is the main eigenvalue of the adjacency matrix $A$; and recover probability $\mu = 1$. Each point represents an airport with a given centrality and spreading capacity, given by the average outbreak size. This average is calculated from 50 simulations of the SIR model starting in each node. The network presents $N = 1,574$ airports and $M = 28,236$ flights. The Pearson correlation coefficient between the percentage of recovered individuals and centrality measures is indicated inside the plots.  Only the network largest component is considered in our analysis. We use the undirected version of the network.}
\label{fig:usair}
\end{figure}

A very different relationship between centrality and node outbreak size can be observed for scale-free networks. For the case of US air transportation network, where each edge represents a connection from one airport to another in 2010, we can see in Figure~\ref{fig:usair} that the eigenvector centrality, degree, and k-core can predict the average outbreak size when the disease starts in each node. The accessibility can also predict, but only for high values of centrality. On the other hand, betweenness centrality is not suitable to predict the disease spreading. 

These two examples show that the choice of the centrality measures to predict the most influential spreaders depends on the network organization. For spatial networks, the random walks accessibility measure is the most suitable, as verified before~\cite{Arruda014}. For scale-free networks, the k-core and eigenvector centrality can determine these nodes very accurately~\cite{Kitsak010}. However, due to the difficulty in the identification of the most influential spreaders for any kind of networks, this problem remains open.

\subsection{Explosive synchronization}

Synchronization is a pervasive process present in many complex systems, such as between neurons in the central nervous system, communication networks, power grids, social interactions, animal behavior, ecosystems and circadian rhythm~\cite{Strogatz04}. These systems are made of self-sustained oscillators that interact forming a network structure. One particularly important type of synchronization is the phase synchronization, which is the process in which oscillators adjust their rhythms according to interaction with their direct neighbors~\cite{Pikovsky03}. A very popular model to describe the evolution of phase oscillators was introduced by Kuramoto~\cite{Rodrigues016}. Each self-sustained oscillator is characterized by its phase $\theta_{i}(t)$, $i=1,...,N$. Pairs of oscillators are connected by the sine of their phase differences. In complex networks, each oscillator $i$ obeys an equation of motion defined as
\begin{equation}\label{eq:kuramoto}
\frac{d\theta_{i}}{dt}= \omega_i + \lambda \sum_{i=1}^{N}A_{ij} \sin(\theta_j - \theta_i), \quad i=1,\ldots,N,
\end{equation}
where $\lambda$ is the coupling strength, $\omega_i$ is the natural frequency of oscillator $i$, and $A_{ij}$ are the elements of the adjacency matrix $A$, so that $A_{ij}=1$ when nodes $i$ and $j$ are connected and $A_{ij}=0$ otherwise. The general Kuramoto model considers a random distribution of the natural frequencies and initial phases according to a specific distribution $g(\omega)$, which is independent of the network structure~\cite{Arenas08}. 

In most of the cases, the frequency distributions are unimodal and symmetric around a mean value $\omega_{0}$~\cite{Arenas08, Rodrigues016}. It is possible to show by mean-field approximation that the critical coupling for the emergence of the synchronous state is given by~\cite{Ichinomiya04:PRE,Restrepo05:PRE}
\begin{equation}
\lambda_{c}^{(0)} = \frac{2}{\pi g(0)} \frac{\left\langle k\right\rangle}{\left\langle k^{2}\right\rangle}.
\label{eq:critical_coupling_s}
\end{equation}
where $g(\omega)$ is symmetric in relation to a single local maximum $\omega_{0}$ (e.g., $\omega_{0}=0$). Thus, the presence of central nodes influences the emergence of the synchronous state, since the more heterogeneous a network, the smaller the critical coupling strength. Indeed, in scale-free networks with $P(k)\sim k^{-\gamma}$, where $\gamma \leq 3$, as $N \rightarrow \infty$ the critical coupling $\lambda_{c}^{(0)}$ goes to zero, since the ratio $\left\langle k\right\rangle / \left\langle k^{2}\right\rangle $ diverges~\cite{Rodrigues016}.

\begin{figure}[!t]
\begin{center}
\subfigure{ \includegraphics[width=.48\linewidth]{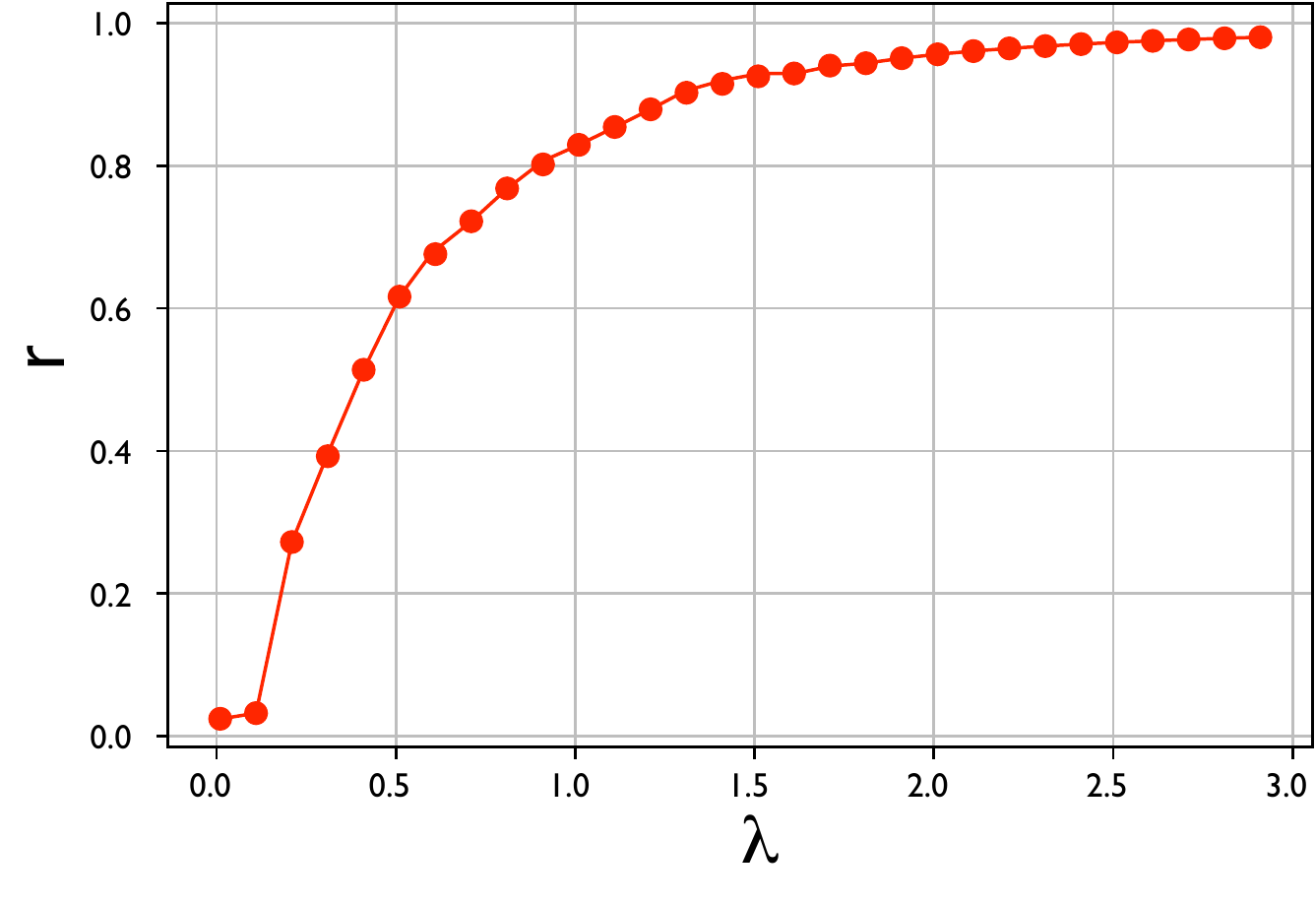}}
\subfigure{ \includegraphics[width=.48\linewidth]{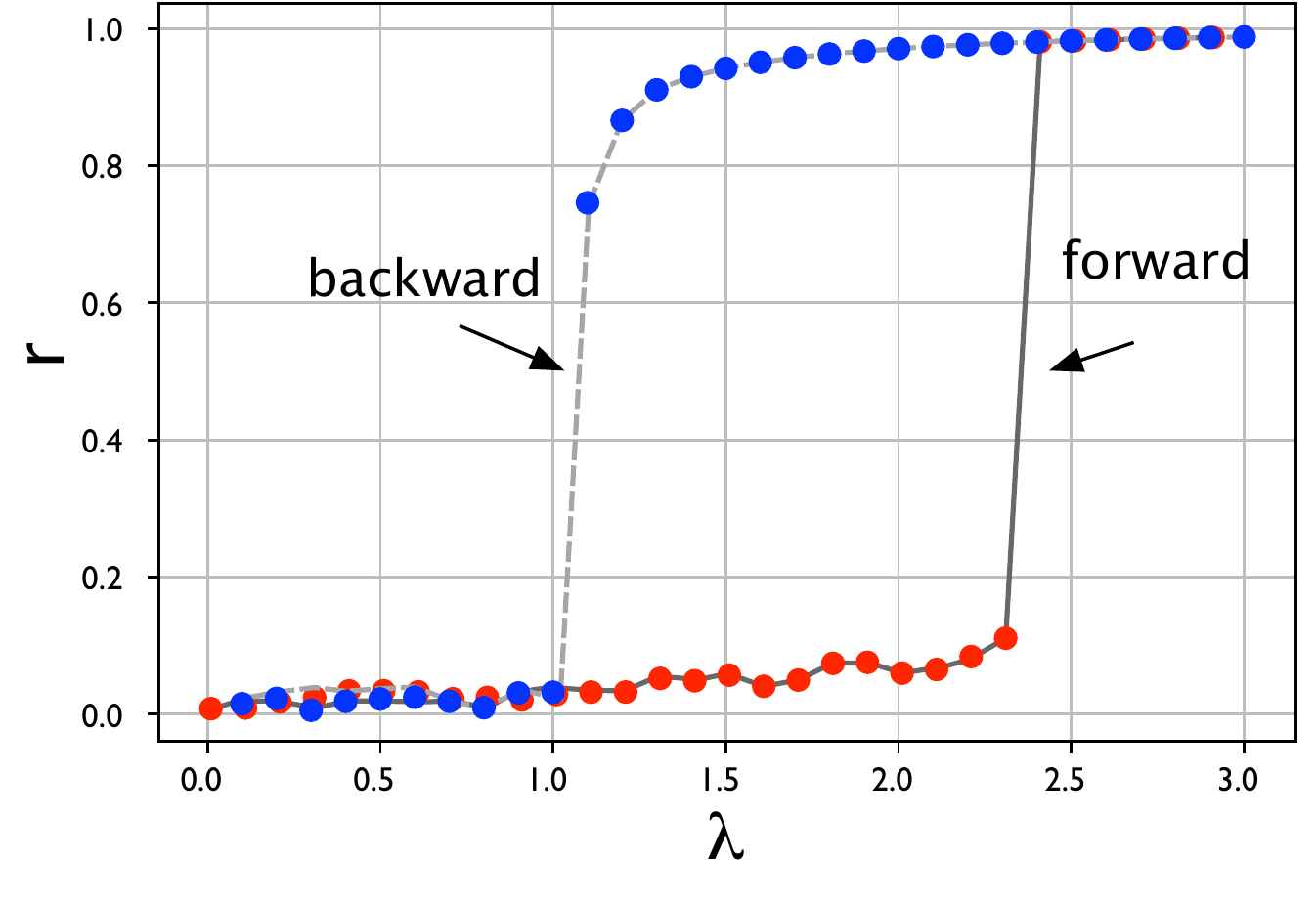}}
\end{center}
\caption{Synchronization diagram $r(\lambda)$ for a Barab\'{a}si-Albert scale-free network with $N=500$ nodes and average degree $\langle k \rangle=4$. (a) Second-order phase transition. The natural frequency of the oscillators is distributed according to a Gaussian distribution with zero mean and unit variance. (b) Explosive synchronization. The natural frequency is equal to the degree, $\omega_i = k_i, \forall i = 1, \ldots, N$. Notice that a hysteretic behavior is observed, indicating a discontinuous transition to synchronization. The forward and backward evolution of $r$ are calculated adiabatically for $\lambda_0, \lambda_0 + \delta \lambda, \ldots,  \lambda_0 + n \delta \lambda$, where we consider  $\delta \lambda = 0.1$.}
\label{fig:sync}
\end{figure}

The level of synchronization is quantified by the macroscopic order parameter, defined as
\begin{equation}
r(t)e^{i\psi(t)} = \frac{1}{N}\sum_{i=1}^{N}e^{i\theta_{i}(t)},
\end{equation}
where the modulus $0 \leq r(t) \leq 1$ and $\psi(t)$ is the average phase of the oscillators. When the natural frequency of each oscillator is independent of the network structure and selected at random from a unimodal and symmetric distribution $g(\omega)$, the order parameter displays a continuous (second-order) phase transition to the synchronous state. This behavior is illustrated in Figure~\ref{fig:sync}(a). On the other hand, when the natural frequency of oscillators is correlated with the number of connection ($\omega_i = k_i$), i.e., when hubs oscillate much faster than the remainder of the nodes, a very different behavior is observed. In this case, a first order phase transition to synchronization occurs in scale-free networks~\cite{Gardenes011}, as shown in Figure~\ref{fig:sync}(b). Thus, the correlation between the natural frequency and the number of connections provides a change in the system behavior. Thus, centrality plays a fundamental role on synchronization in uncorrelated networks.

Further works verified that the correlation between the natural frequency and the number of connections is not a unique condition for the emergence of explosive synchronization. The key mechanism behind explosive synchronization is the existence of sufficient large frequency mismatches between nodes and their neighbors, what occurs in non-correlated and disassortative scale-free networks~\cite{Rodrigues016}. However, the effect of centrality on synchronization is a key feature that deservers a deeper investigation.

\section{Applications}

Network centrality has been used to characterize complex systems in many areas, ranging from climate to biological interactions. Here we present some applications of centrality measures to characterize the structure and function of complex systems. One of the first applications of network centrality was in the prediction of protein lethality. Jeong et al. verified that the most connected proteins of the yeast \textit{S. cerevisiae} are more likely essential than low connected proteins~\cite{Jeong01}. A similar observation was verified in genetic networks, where the P53 gene, which is a tumor-suppressor, is at the center of the network. Therefore, the disruption of p53 has severe consequences to keep cells under control~\cite{Vogelstein00}. 

In medicine, human disease genes have a tendency to encode hubs in the interactome, which may be a consequence of the evolutionary process~\cite{Goh07}. For instance, squamous cell lung cancer genes are the most central in human protein interaction maps, sharing similar topological features as essential proteins~\cite{Wachi05}. Moreover, eigenvector and degree centrality were used to identify genes related to prostate cancer~\cite{Ozg08} with high accuracy. This relation between centrality and functions in biological networks has allowed disease gene predictions and can help to develop individualized therapies and potential cures for many diseases~\cite{Goh07}.

In cortical networks, centrality also plays a fundamental role. Centrality measures are fundamental to quantify the large-scale organization of cortical networks. For instance, by using fMRI data Achard et al.~\cite{Achard06} verified that the human brain is dominated by a neocortical core of highly connected hubs, that have long-distance connections to other regions. These hubs are associated with network robustness since cortical networks revealed to be more resilient to a targeted attack on their hubs than a comparable scale-free network. The authors verified that regional hubs represent large-scale neocortical-thalamic circuits, which are related to human cognition and consciousness. Therefore, central nodes play a fundamental role in brain functioning. In addition to the degree, different centrality measure captures different aspects of connectivity, as verified by Zuo et al.~\cite{Zuo011}. The authors considered data from 1003 subjects and verified age-related decreases in degree centrality within precuneus and posterior cingulate regions, whereas the eigenvector centrality remains constant with age. The difference in centrality level was also verified in the case of child-onset schizophrenia. Arruda et al.~\cite{Arruda014} observed that cortical networks of schizophrenic subjects present higher values of the variance of the closeness centrality and accessibility, but small values of average k-core. By considering these measures, they verified that it is possible to perform reliable automatic diagnostics in patients with schizophrenia with a sensitivity of 90\% and specificity of 74\%. Among 54 network measures extracted from the cortical networks, the authors verified that only four (three related to centrality) of them differ schizophrenic and healthy subjects. 
 
In climate data analysis, regions presenting the largest values of betweenness centrality are related to global surface ocean currents~\cite{Donges09}. The authors verified that the nodes with the highest betweenness centrality form the backbone of the network, that has an essential role on the oceanic surface circulation~\cite{Donges09}. In world trade networks, central countries hold many trade partners and the richest nations have a very intense trade relationship~\cite{Fagiolo09}. In the worldwide air transportation network, the most connected cities are not necessarily the most central~\cite{Guimera05}, which occurs because of the multi-community structure of the network. Several applications of network theory on the analysis of complex systems can be found in~\cite{Costa011}.

\section{Perspectives}


Centrality can be defined in terms of different concepts, like random walks and connectivity. The choice of the most suitable measure to characterize a complex network is application dependent, as we have verified in the case of the identification of influential spreaders. This lack of generality opens a new possibility for proposing new centrality measures. Moreover, centrality measures are little explored in the characterization of many dynamical processes, like cooperation and opinion formation. For instance, in cooperative processes, only a few works verified the importance of centrality for improving learning tasks~\cite{Reia017}.

Moreover, new kinds of networks have been studied in the last years. Many complex systems exhibit more than one type of interaction, like in social networks, where the same set of users share information in Twitter and Google+. In this case, the system's organization can be represented by a multilayer network~\cite{Kivel014, Boccaletti014}, in which each layer represents one type of interaction. In these networks, the definition of centrality has to consider the edges of each layer and the connections between layers. Recently, some versions of network centrality indices have been proposed for quantification of node importance~\cite{Domenico013, Kivel014, Boccaletti014}. However, there are many possibilities of research on the applications of these new metrics for characterization of complex systems, like in ecology, economy and social networks, as well as the study of the influence of central nodes on dynamical processes in networks.

Temporal networks are also new generalizations of static complex networks~\cite{Holme012}. In this case, the connections between nodes change with time, which affects the system behavior, as in epidemic spreading in social networks whose contacts occurs in different time intervals. Centrality measures have been adapted to these networks. However, they have mostly been generalizations of static network measures and more improvements and further analyses are needed~\cite{Holme012}. These studies are very important for a better understanding of the evolution of complex systems and the role of centrality for dynamical processes, like epidemic spreading and synchronization in networks whose connections are not static.

Although centrality is a key concept in network theory, these generalizations and applications in many fields should help us to understand the structure, function, and dynamics of complex systems. A better understanding of the relation between centrality metrics and dynamics can provide efficient tools to control and forecast different dynamical processes.

\section*{Acknowledgments}

The author thanks Jos\'{e} Fernando Fontanari for useful comments. This work was funded in part by CNPq (grant 305940/2010-4), and FAPESP (grants 2016/25682-5 and grants 2013/07375-0).

\bibliographystyle{plain}
\bibliography{references}

\end{document}